\documentclass[letter]{aa}
\usepackage{textgreek}
\usepackage{graphicx}
\usepackage{comment}
\usepackage{mathtools}
\usepackage{amsmath}
\usepackage{fancyref}
\usepackage{longtable}
\usepackage{rotating}
\usepackage{pdflscape}
\usepackage{enumerate}
\usepackage{xtab,booktabs}
\usepackage{textcomp}
\usepackage[citecolor=blue, urlcolor=blue, linkcolor=blue, colorlinks=true]{hyperref}
\usepackage{txfonts}
\usepackage{soul}
\usepackage{natbib}
\usepackage{subcaption}
\usepackage{bm}

\usepackage{afterpage}
\usepackage{caption}
\usepackage{float}
\usepackage{lipsum}% dummy text
\usepackage{ulem}
\usepackage{booktabs}

\usepackage{ragged2e}

\usepackage{tikz,xcolor,hyperref}

\definecolor{lime}{HTML}{A6CE39}
\DeclareRobustCommand{\orcidicon}{%
	\hspace{-1.5mm}
	\begin{tikzpicture}
	\draw[lime, fill=lime] (0,0) 
	circle [radius=0.16] 
	node[white] {{\fontfamily{qag}\selectfont \tiny ID}};
	\draw[white, fill=white] (-0.0625,0.095) 
	circle [radius=0.007];
	\end{tikzpicture}
	\hspace{-2.5mm}
}
\foreach \x in {A, ..., Z}{%
	\expandafter\xdef\csname orcid\x\endcsname{\noexpand\href{https://orcid.org/\csname orcidauthor\x\endcsname}{\noexpand\orcidicon}}
}

\foreach \x in {a, ..., z}{%
	\expandafter\xdef\csname orcid\x\endcsname{\noexpand\href{https://orcid.org/\csname orcidauthor\x\endcsname}{\noexpand\orcidicon}}
}

\begin{document}

   \title{Peering through the disc of HD~98800~BaBb}
   \subtitle{Precise timing predictions for the HD~98800~AaAb occultation}

  \author{S. Z\'u\~niga-Fern\'andez 
          \orcidA{}\inst{\ref{astro_liege}} \thanks{Corresponding author: \href{mailto:sgzuniga@uliege.be}{sgzuniga@uliege.be}}
          \and A. Bayo 
          \orcidC{}\inst{2}
          \and J. Olofsson
          \orcidB{}\inst{2}
          \and J. Ehrhardt
          \orcidD{}\inst{2}
          \and \'A. Ribas
          \orcidE{}\inst{3}
          }
   \institute{
  Astrobiology Research Unit, Universit\'e de Li\`ege, All\'ee du 6 Ao\^ut 19C, 4000 Li\`ege, Belgium \label{astro_liege}
   \and
    European Southern Observatory, Karl-Schwarzschild-Str. 2, 85748, Garching bei M\"unchen, Germany
    \and
    Astronomy Unit, Department of Physics and Astronomy, Queen Mary University of London, Mile End Road, London E1 4NS, United Kingdom
    }
    
    \date{Received 11 June 2026 / Accepted 11 July 2026}

\abstract 
{The young hierarchical quadruple system HD 98800 is composed of the tight binaries AaAb and BaBb on a wide, highly inclined outer orbit. The B subsystem hosts a circumbinary disc in a polar configuration, and the geometry of the system offers a rare opportunity to observe the passage of the disc around BaBb in front of AaAb.}
{We update the overall orbital solution to offer precise time windows of the main occultation features by combining the orbit based on the most recent knowledge of the disc structure.}
{We combined new and published radial velocity measurements and multi-wavelength astrometry for the outer AB orbit and the two inner subsystems, AaAb and BaBb, in a joint orbital fit.}
{The revised solution is consistent with previous dynamical masses and improves the outer orbit, reducing the uncertainties in the period and periastron epoch by about a factor of two. It also narrows the predicted crossing-phase windows to 5--15 days at the 1$\sigma$ level, improving the timing of ingress, egress, and the first cavity-crossing predictions.}
{These results provide a more accurate timing framework for future observations of the occultation, although the predicted epochs remain model-dependent because of uncertainties in the disc structure.}


\maketitle

\section{Introduction}
\label{sec:intro} 
Multiple systems provide a direct laboratory for studying the co-evolution of stellar orbits and discs during the pre-main-sequence phase \citep{Czekala2021,Kraus2020}. The system HD~98800 is particularly interesting because it is a young quadruple composed of two tight binaries, AaAb and BaBb, on a wide, highly inclined orbit, with the B subsystem hosting a circumbinary disc \citep{Boden2005,Kennedy2019}. The estimated combined spectral types are K5\,V for AaAb and K7\,V for BaBb \citep{Soderblom1998}. As a member of the TW Hydrae association \citep{Torres2006}, its age of $\sim 10$ Myr makes the survival of the disc in this dynamically complex environment especially remarkable. \cite{Ribas2018} suggested that tidal interactions with the inner and outer binaries may slow the viscous evolution of the disc, and \cite{Ronco2021} showed with 1D numerical simulations that the A subsystem can significantly extend the gas dispersal timescale of the B subsystem. \cite{GiupponeCuello2019} and \cite{Ceppi2024} showed that the polar disc configuration can arise as a stable outcome of dynamical interactions in multiple systems.

The HD~98800 architecture is especially valuable because the inner binaries and the outer orbit evolve on timescales that are short enough for their orbital configuration to be constrained and monitored with current observations. \citet{SZF2021} (SZF21 hereafter) obtained the first full orbital solution for AaAb, refined the BaBb orbit, and revisited the AB outer orbit by deriving dynamical masses and showing that the system is expected to be stable over long timescales. Together with a simplified disc-opacity model, these results suggested that AaAb is expected to be occulted by the disc around BaBb at some point in 2026 \citep{Kennedy2019,SZF2021}.

The timing of the predicted occultation is particularly important because this event offers a key observational opportunity. Synthetic light curves based on a physically motivated disc model show that the event will probably last several years, which is primarily set by the orbital geometry, while its detailed morphology depends on the disc dust mass, gas content, and viscosity \citep{Faruqi2025}. \citet{Faruqi2025} further showed that the light-curve shape can encode the radial extent of the disc, its asymmetries, and its truncation by the wide orbit.

In this Letter, we refine the orbital solution using new spectroscopy and astrometry, and we revise the timing of the forthcoming occultation of AaAb by the polar circumbinary disc. By combining the disc geometry adopted by \citet{Faruqi2025} with the updated outer orbit, we refine the predicted occultation window, in particular, for the ingress and egress phases.

\begin{table*}
\small
\centering
\caption{Orbital parameters for the HD\,98800 AaAb and BaBb binaries.}
\label{tab:orbit_AaAb_BaBb}
\renewcommand{\arraystretch}{1.15}
\begin{tabular}{lcccc}
\toprule
 & \multicolumn{2}{c}{BaBb} & \multicolumn{2}{c}{AaAb} \\
Orbital parameters & SZF21 & This work & SZF21 & This work \\
\midrule
\multicolumn{5}{l}{Fitted parameters} \\
\midrule
Period (d)      & $314.86 \pm 0.02$ & $314.87848 \pm 0.00015$ & $264.51 \pm 0.02$ & $264.507 \pm 0.020$ \\
$T_{0}$ (MJD)      & $48707.5 \pm 0.2$ & $48708.70 \pm 0.21$ & $48742.5 \pm 0.8$ & $48742.46 \pm 0.79$ \\
$e$                & $0.805 \pm 0.005$ & $0.7819 \pm 0.0049$ & $0.4808 \pm 0.0008$ & $0.48082 \pm 0.00085$ \\
$\omega$ ($^\circ$) & $104.5 \pm 0.3$ & $104.86 \pm 0.34$ & $68.7 \pm 0.1$ & $68.70 \pm 0.10$ \\
$\Omega$ ($^\circ$) & $342.7 \pm 0.4$ & $344.02 \pm 0.32$ & $170.2 \pm 0.1$ & $170.17 \pm 0.15$ \\
$i$ ($^\circ$)      & $66.3 \pm 0.5$ & $65.43 \pm 0.31$ & $135.6 \pm 0.1$ & $135.59 \pm 0.12$ \\
$a$ (mas)           & $22.2 \pm 0.4$ & $22.04 \pm 0.23$ & $19.03 \pm 0.01$ & $19.036 \pm 0.015$ \\
$K_{1}$ (km\,s$^{-1}$)  & $24.0 \pm 0.3$ & $22.63 \pm 0.29$ & $6.7 \pm 0.2$ & $6.79 \pm 0.14$ \\
$K_{2}$ (km\,s$^{-1}$)  & $29.9 \pm 0.6$ & $28.11 \pm 0.38$ & \dots & \dots \\
$\gamma_{~\mathrm{TO95}}$ (km\,s$^{-1}$)  & $5.6 \pm 0.1$ & $5.65 \pm 0.14$ & $12.8 \pm 0.1$ & $12.840 \pm 0.098$ \\
$\gamma_{~\mathrm{ELODIE}}$ (km\,s$^{-1}$) & $3.4 \pm 0.7$ & $3.85 \pm 0.40$ & $12.1 \pm 0.5$ & $12.09 \pm 0.46$ \\
$\gamma_{~\mathrm{CTIO}}$ (km\,s$^{-1}$)   & $6.4 \pm 0.4$ & $6.61 \pm 0.29$ & $11.9 \pm 0.2$ & $11.86 \pm 0.18$ \\
$\gamma_{~\mathrm{FEROS}}$ (km\,s$^{-1}$)   & \dots & \dots & $14.3 \pm 0.3$ & $14.33 \pm 0.31$ \\
$\gamma_{~\mathrm{FEROS2}}$ (km\,s$^{-1}$)  & \dots & \dots & $12 \pm 1$ & $11.9 \pm 1.1$ \\
$\gamma_{~\mathrm{Mercator}}$ (km\,s$^{-1}$) & \dots  & $4.73 \pm 0.27$ & \dots & $12.86 \pm 0.17$ \\
$\gamma_{~\mathrm{HARPS}}$ (km\,s$^{-1}$)   & \dots & $7.06 \pm 0.50$ & \dots & $11.83 \pm 0.77$ \\
$\gamma_{~\mathrm{NIRPS}}$ (km\,s$^{-1}$)   & \dots & $7.080 \pm 0.097$ & \dots & $11.54 \pm 0.68$ \\
$f_{~\mathrm{K,Bb}}/f_{~\mathrm{K,Ba}}$   & $0.76 \pm 0.08$ & $0.65 \pm 0.10$ & \dots & \dots \\
\midrule
\multicolumn{5}{l}{Derived parameters} \\
\midrule
$\pi$ (mas)           & $22.0 \pm 0.6$ & $21.90 \pm 0.35$ & $22.0 \pm 0.6$ & $21.90 \pm 0.35$ \\
$M_{1}$ ($M_{\odot}$) & $0.77 \pm 0.04$ & $0.760 \pm 0.020$ & $0.93 \pm 0.09$ & $0.956 \pm 0.059$ \\
$M_{2}$ ($M_{\odot}$) & $0.62 \pm 0.02$ & $0.612 \pm 0.013$ & $0.29 \pm 0.02$ & $0.298 \pm 0.013$ \\
$d$ (pc)              & $45 \pm 1$ & $45.66 \pm 0.71$ & \dots & $45.66 \pm 0.71$ \\
$a$ (au)              & $1.01 \pm 0.01$ & $1.0064 \pm 0.0075$ & $0.86 \pm 0.02$ & $0.870 \pm 0.016$ \\
\bottomrule
\end{tabular}
\end{table*}
%\LEt{***please provide the spelled-out names of all instruments and surveys at first occurrence in the main text so that the spelled-out name is part of the sentence and the abbreviation or acronym is given in parentheses. This applies throughout. Please check and amend as required. I'll not highlight this again to avoid cluttering the ms***}
\section{Observations}\label{sec:observations}
The dataset we analysed consists of the measurements used in SZF21, together with four new radial velocity epochs for AaAb and BaBb and new astrometric measurements of the BaBb and AB subsystems. The radial velocities were obtained from High Accuracy Radial velocity Planet Searcher \citep[HARPS,][]{Mayor2003} and Near-Infrared Planet Searcher \citep[NIRPS,][]{Bouchy2025} observations, which are the two high-resolution spectrographs at the ESO 3.6\,m telescope. We also included published Hermes/Mercator radial velocity data \citep{Merle2024}. The astrometric data include a new near-infrared VLTI/PIONIER \citep{LeBouquin2011} measurement of BaBb, speckle-interferometric observations obtained with the Zorro camera on Gemini South \citep{Scott2021} by \citet{Mendez2025}, and two epochs of Very Large Array \citep[VLA,][]{Thompson2017} astrometry obtained from \cite{Ribas2018} and \cite{Ribas2026}. These data extend the time coverage of the system and provide additional constraints for the orbital analysis. The observing logs, measurement tables, and the methods for obtaining the new radial velocity and VLA astrometric data are given in the Appendix \ref{sec:appendix_observations}.

\section{Orbital fitting}\label{sec:orbital_fit}
The updated dataset was modelled with the same joint framework for astrometry and radial velocity as was used by SZF21. It combines measurements from the different instruments through a single orbital solution with instrument-specific velocity zero points. We used the \texttt{exoplanet} package \citep{exoplanet:exoplanet}, which extends the PyMC3 framework \citep{exoplanet:pymc3} and provides the custom functions and distributions required for orbital fitting. As in SZF21, the orbital parameters were sampled with broad priors and estimated from the posterior distributions, taking the median as the best-fit value and the larger of the upper and lower 16th-84th percentile deviations as the quoted uncertainty. Prior distributions and orbital solution plots are provided in Appendix~\ref{sec:apendix_orbits}. All predictions for the orbital fit code, posterior distributions, corner plots, and orbital motions are also available in a public repository for reproducibility and future comparisons with the observed event\footnote{\url{https://github.com/szunigaf/HD98800-orbit}}.

\subsection{BaBb}
The orbital fit of BaBb was performed following the same procedure for joint astrometry and radial velocity as in SZF21. It combines the available relative astrometry, near-infrared squared-visibility ($V^2$) data from the Keck Interferometer (KI) and radial velocities in a single orbital solution. Compared with the previous analysis, the updated fit also includes the new observations presented here. Following standard practice for KI archival data, the pipeline uncertainties were found to be underestimated because of instrumental systematics and atmospheric phase jitter \citep{Colavita2010,Mennesson2014}. To account for this effect, a jitter term was added in quadrature to the pipeline-generated $V^2$ errors. Compared with SZF21, the revised fit yields substantially smaller uncertainties for the period (from 0.02 d to 0.0001 d) and parallax (from 22.0 $\pm$ 0.6 mas to 21.9 $\pm$ 0.4 mas), improving the distance and absolute scale. The semi-major axis is slightly smaller and better constrained (22.1 $\pm$ 0.2 mas vs.\ 22.2 $\pm$ 0.4 mas), and the component masses are refined to 0.76 $\pm$ 0.02 $M_{\odot}$ and 0.615 $\pm$ 0.01 $M_{\odot}$ (see Table~\ref{tab:orbit_AaAb_BaBb}). The eccentricity and orientation angles remain consistent within the uncertainties.

\subsection{AaAb}
Compared with SZF21, the updated AaAb solution is largely consistent with the earlier results, but provides modest improvements in several derived quantities (see Table~\ref{tab:orbit_AaAb_BaBb}). In SZF21, the AaAb orbit was obtained for the first time by combining PIONIER relative astrometry with radial velocities in a joint fit, which also tied the dynamical masses of the AaAb subsystem to the parallax inferred from the BaBb solution. The orbital period, epoch, and geometry ($e$, $\omega$, $\Omega$, $i$) remain unchanged here within the uncertainties, while the angular semi-major axis and velocity amplitude are slightly tightened ($a = 19.04 \pm 0.01$ mas; $K_{1} = 6.8 \pm 0.1\ \mathrm{km\,s^{-1}}$). The revised parallax from the BaBb fit ($\pi = 21.9 \pm 0.4$ mas) yields slightly higher component masses, which are now $M_{~\mathrm{Aa}} = 0.96 \pm 0.06\,M_{\odot}$ and $M_{~\mathrm{Ab}} = 0.30 \pm 0.01\,M_{\odot}$, and a marginally larger physical semi-major axis ($a = 0.87 \pm 0.02$ au).

\subsection{AB}
The outer AB orbit was fitted using the same strategy for joint astrometry and radial velocity as in SZF21. The visual AB astrometry was combined with the systemic radial velocities of the A and B subsystems in a single orbital solution. We note that the astrometric measurements taken before 1991 have unknown uncertainties; for these points, we therefore adopted two typical error values within the range reported by USNO astrometry measurements \citep{Torres1999}, and we defined large ($\sigma \sim 0.1$) and small ($\sigma \sim 0.02$) uncertainty cases (solutions I and II, respectively, in Table~\ref{tab:orbit_AB}). Compared with SZF21, our revised solutions reduce the parameter uncertainties while remaining consistent with the previous work. The eccentricity and orientation are consistent with SZF21 within the errors, while $K_A$, $K_B$, the systemic velocity, and the component masses are now much better constrained ($M_{~\mathrm{AaAb}} = 1.26 \pm 0.08\,M_\odot$; $M_{~\mathrm{BaBb}} = 1.31 \pm 0.08\,M_\odot$) than in our previous work. The difference of $\sim$2.2 years in $T_0$ between solutions I and II (and the corresponding 20-year spread in $P$) translates into only a modest positional uncertainty for the A and B components over the next 10 years. With the orbital period of $\sim$230 years and angular semi-major axis $a\sim1.1''$, the mean orbital motion is $\sim30$ mas\,yr$^{-1}$. This means that a 2-year $T_0$ offset produces a separation uncertainty of only $\sim60$ mas over 5 years, which is small compared to the full orbit size and current astrometric precision.

\begin{table}
\small
\centering
\caption{Orbital parameters for the HD\,98800 AB system.}
\label{tab:orbit_AB}
\renewcommand{\arraystretch}{1.15}
\begin{tabular}{lccc}
\toprule
Fitted parameters & SZF21 & Solution I & Solution II \\
\midrule 
Period (yr)                  & $230 \pm 20$    & 212$\pm$11 & 237$\pm$12 \\
$T_0$ (yr)                   & $2023 \pm 1$    & 2024.11$\pm$0.79 & 2021.63$\pm$0.65 \\
$e$                          & $0.46 \pm 0.02$ & 0.437$\pm$0.014 & 0.460$\pm$0.015 \\
$\omega_A$ ($^\circ$)        & $65 \pm 5$      & 69.1$\pm$3.5 & 58.8$\pm$2.6 \\
$\Omega$ ($^\circ$)          & $184.5 \pm 0.1$ & 184.47$\pm$0.10 & 184.639$\pm$0.095 \\
$i$ ($^\circ$)              & $88.1 \pm 0.1$  & 87.850$\pm$0.082 & 88.074$\pm$0.064 \\
$\gamma_{~\mathrm{AB}}$ (km s$^{-1}$) & $8.7 \pm 0.7$   & 9.15$\pm$0.20 & 9.16$\pm$0.22 \\
$M_{~\mathrm{AaAb}}$ ($M_\odot$)           & $1.1 \pm 0.3$   & 1.262$\pm$0.075 & 1.280$\pm$0.085 \\
$M_{~\mathrm{BaBb}}$ ($M_\odot$)           & $1.4 \pm 0.3$   & 1.306$\pm$0.085 & 1.315$\pm$0.086 \\
$\pi$ (mas)                 & $22.2 \pm 0.5$  & 21.89$\pm$0.31 & 22.27$\pm$0.28 \\
\midrule
Derived parameters & & & \\
\midrule 
$K_{~\mathrm{AaAb}}$ (km s$^{-1}$)         & $4.2 \pm 0.8$   & 3.86$\pm$0.23 & 3.78$\pm$0.20 \\
$K_{~\mathrm{BaBb}}$ (km s$^{-1}$)         & $3.2 \pm 0.8$   & 3.73$\pm$0.20 & 3.67$\pm$0.20 \\
$a$ ($^{\prime\prime}$)      & $1.13 \pm 0.08$ & 1.067$\pm$0.046 & 1.172$\pm$0.044 \\
$a$ (au)                     & $51 \pm 3$      & 48.8$\pm$1.9 & 52.6$\pm$2.0 \\
$d$ (pc)                     & $45 \pm 1$      & 45.68$\pm$0.70 & 44.91$\pm$0.57 \\
\bottomrule
\end{tabular}
\end{table}

\section{Predicted occultation epochs}\label{sec:results}

\begin{figure*}[t]
\centering
   \includegraphics[width=0.99\textwidth]{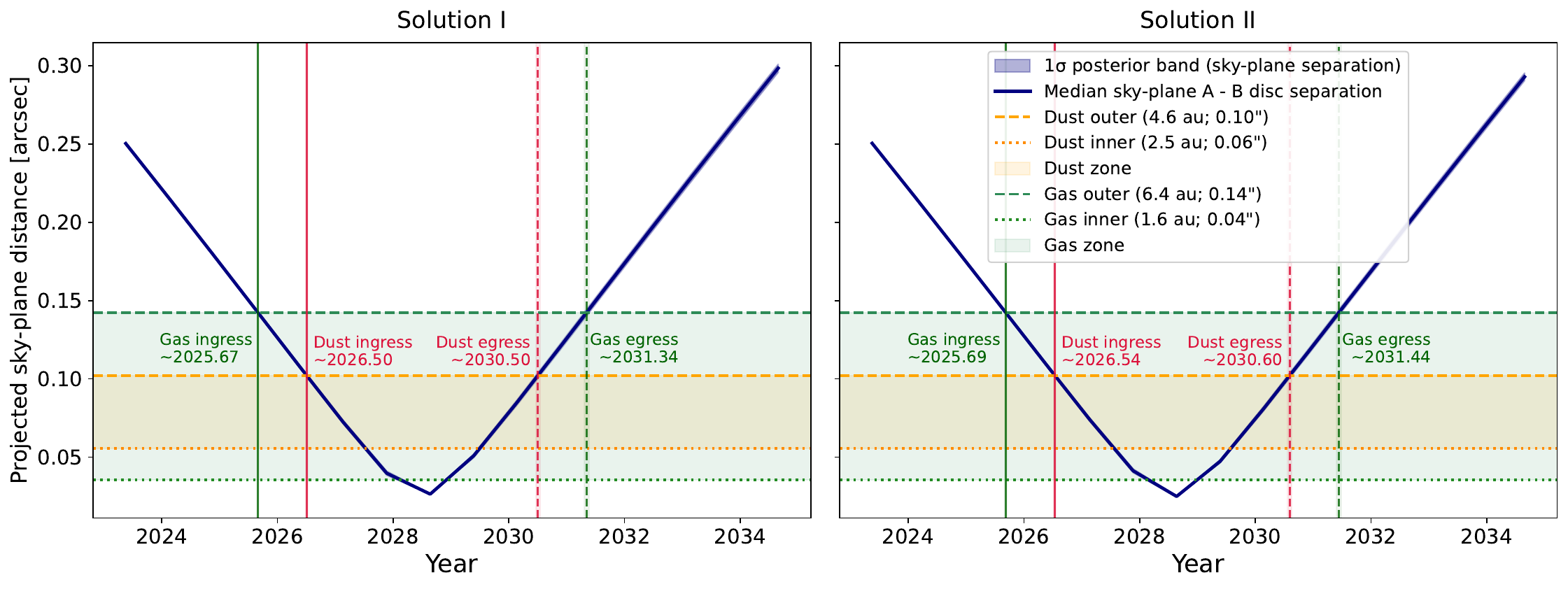}
   \vspace{-0.3cm}
\caption{Projected sky-plane separation of HD~98800\,AaAb relative to the HD 98800\,BaBb disc as a function of time, compared with the disc geometry adopted in \cite{Faruqi2025}. The model is used to identify the onset of the occultation of HD~98800\,AaAb by the circumbinary disc around HD~98800\,BaBb. The shaded region marks the predicted occultation windows for the dust and gas edges, and the vertical line indicates the median ingress and egress epochs.}
\label{Fig:A_occulation}
\end{figure*}

The revised AB orbital solutions improve the prediction of the forthcoming occultation of AaAb by the circumbinary disc around BaBb (Table~\ref{tab:occultation_window}; Fig.~\ref{Fig:A_occulation}) while remaining fully consistent with SZF21 within the uncertainties. In Fig.~\ref{Fig:AB_disc_model} we overplot a sketch of the mid-March disc position in the sky plane, based on the disc geometry adopted by \cite{Faruqi2025} for the two solutions. The astrometric error bars are shown only in Fig.~\ref{Fig:AB_orbit_sepPA}. Relative to SZF21, the updated fits reduce the uncertainties on the outer orbital period and periastron epoch by about a factor of two, narrowing the range of the estimated sky positions of A relative to B over the interval $2023-2031$. To translate the orbital posterior into occultation epochs, we projected the adopted inclined disc geometry onto the sky plane and compared it with the sky-plane orbit of A relative to B (see Appendix~\ref{sec:app_occultation}). We adopted the disc geometry inferred from the ALMA-based models of \citet{Kennedy2019} and used by \citet{Faruqi2025}, namely an inclination of 26\degr, a major-axis position angle of 15.6\degr, a dust annulus extending from 2.5 to 4.6 au, and a more extended gas disc from 1.6 to 6.4 au. Ingress and egress epochs were then identified from the times at which the sky-projected orbit crossed the projected dust- and gas-disc boundaries. We evaluated these crossings for the 16th, 50th, and 84th percentile orbital tracks, which provided a median epoch and a formal 1$\sigma$ interval for each disc-crossing phase. The quoted dates in Table~\ref{tab:occultation_window} correspond to the early ($-1\sigma$), median, and late ($+1\sigma$) limits of the predicted sky-plane crossing window. The resulting windows are narrower than in SZF21, in some cases as short as 5--15 days, and they therefore offer a more precise basis for planning time-critical observations. These values, however, should be regarded as model-dependent estimates, since they only reflect the uncertainty from the orbital posterior and assume fixed disc radii and orientation. The true ingress and egress times may shift if the disc structure differs from the adopted ALMA-based geometry. If the outer dust annulus is optically thin and the first measurable dimming is produced by the optically thick inner edge, the photometric ingress could occur later than the geometric dust-edge crossing. On the other hand, if low-density dust extends beyond the nominal disc edge, the occultation could begin earlier than predicted. The estimated sequence of gas and dust crossings provides a useful observational framework for future monitoring of the event. In particular, multi-band photometric and spectroscopic observations obtained across ingress, the cavity crossings, the main occultation phase, and egress are expected to help us to distinguish the radial extent and opacity structure of the obscuring material. The cavity itself may be especially informative, since ingress and egress through this region can probe disc dispersal and disc–binary interaction in addition to the outer-disc truncation. These observations can also test whether the occulting disc follows the simple axisymmetric geometry adopted in this work.

\begin{table}
\small
\caption{Predicted time windows for the occultation of AaAb by the
BaBb disc in HD~98800.}
\renewcommand{\arraystretch}{1.15}
\centering
\begin{tabular}{lccc}
\toprule
Crossing phase & Early ($-1\sigma$) & Median & Late ($+1\sigma$) \\
\midrule
\multicolumn{4}{c}{Solution I} \\
\midrule
Gas outer ingress & 26 Aug 2025 & 1 Sep 2025 & 6 Sep 2025 \\
Dust outer ingress & 26 Jun 2026 & 3 Jul 2026 & 11 Jul 2026 \\
Dust inner egress & 1 Jul 2027 & 11 Jul 2027 & 22 Jul 2027 \\
Gas inner egress & 26 Jan 2028 & 16 Feb 2028 & 5 Mar 2028 \\
Gas inner ingress & 28 Nov 2028 & 2 Dec 2028 & 6 Dec 2028 \\
Dust inner ingress & 20 Jun 2029 & 30 Jun 2029 & 11 Jul 2029 \\
Dust outer egress & 20 Jun 2030 & 3 Jul 2030 & 18 Jul 2030 \\
Gas outer egress & 20 Apr 2031 & 5 May 2031 & 21 May 2031 \\
\midrule
\multicolumn{4}{c}{Solution II} \\
\midrule
Gas outer ingress & 4 Sep 2025 & 9 Sep 2025 & 14 Sep 2025 \\
Dust outer ingress & 8 Jul 2026 & 15 Jul 2026 & 22 Jul 2026 \\
Dust inner egress & 15 Jul 2027 & 25 Jul 2027 & 4 Aug 2027 \\
Gas inner egress & 10 Feb 2028 & 26 Feb 2028 & 11 Mar 2028 \\
Gas inner ingress & 26 Dec 2028 & 29 Dec 2028 & 3 Jan 2029 \\
Dust inner ingress & 21 Jul 2029 & 30 Jul 2029 & 8 Aug 2029 \\
Dust outer egress & 25 Jul 2030 & 6 Aug 2030 & 19 Aug 2030 \\
Gas outer egress & 28 May 2031 & 11 Jun 2031 & 25 Jun 2031 \\
\bottomrule
\end{tabular}
\label{tab:occultation_window}
\end{table}

\newpage

\begin{acknowledgements}
      AB and JE acknowledge support from the Deutsche Forschungsgemeinschaft under Germany's Excellence Strategy – EXC 2094 – 390783311. AR has received funding from the Royal Society through a University Research Fellowship grant number URF\textbackslash R1\textbackslash 241791. Based on observations collected at the European Southern Observatory under ESO programme 105.20JX and 111.255C. This research has made use of the Washington Double Star Catalog maintained at the U.S. Naval Observatory. This research made use of \textsf{exoplanet} \citep{exoplanet:exoplanet} and its dependencies \citep{exoplanet:astropy13, exoplanet:astropy18, exoplanet:pymc3, exoplanet:theano}. This research has made use of the Jean-Marie Mariotti Center OiDB service available at \url{http://oidb.jmmc.fr}. We thank the Editor and the referee for their helpful comments and suggestions.
\end{acknowledgements}

\vspace{0.2cm}
\bibliography{biblio}

\begin{thebibliography}{32}
\expandafter\ifx\csname natexlab\endcsname\relax\def\natexlab#1{#1}\fi

\bibitem[{{Astropy Collaboration} {et~al.}(2018){Astropy Collaboration}, {Price-Whelan}, {Sip{\H o}cz}, {G{\"u}nther}, {Lim}, {Crawford}, {Conseil}, {Shupe}, {Craig}, {Dencheva}, {Ginsburg}, {VanderPlas}, {Bradley}, {P{\'e}rez-Su{\'a}rez}, {de Val-Borro}, {Aldcroft}, {Cruz}, {Robitaille}, {Tollerud}, {Ardelean}, {Babej}, {Bach}, {Bachetti}, {Bakanov}, {Bamford}, {Barentsen}, {Barmby}, {Baumbach}, {Berry}, {Biscani}, {Boquien}, {Bostroem}, {Bouma}, {Brammer}, {Bray}, {Breytenbach}, {Buddelmeijer}, {Burke}, {Calderone}, {Cano Rodr{\'{\i}}guez}, {Cara}, {Cardoso}, {Cheedella}, {Copin}, {Corrales}, {Crichton}, {D'Avella}, {Deil}, {Depagne}, {Dietrich}, {Donath}, {Droettboom}, {Earl}, {Erben}, {Fabbro}, {Ferreira}, {Finethy}, {Fox}, {Garrison}, {Gibbons}, {Goldstein}, {Gommers}, {Greco}, {Greenfield}, {Groener}, {Grollier}, {Hagen}, {Hirst}, {Homeier}, {Horton}, {Hosseinzadeh}, {Hu}, {Hunkeler}, {Ivezi{\'c}}, {Jain}, {Jenness}, {Kanarek}, {Kendrew}, {Kern}, {Kerzendorf}, {Khvalko}, {King}, {Kirkby}, {Kulkarni},
  {Kumar}, {Lee}, {Lenz}, {Littlefair}, {Ma}, {Macleod}, {Mastropietro}, {McCully}, {Montagnac}, {Morris}, {Mueller}, {Mumford}, {Muna}, {Murphy}, {Nelson}, {Nguyen}, {Ninan}, {N{\"o}the}, {Ogaz}, {Oh}, {Parejko}, {Parley}, {Pascual}, {Patil}, {Patil}, {Plunkett}, {Prochaska}, {Rastogi}, {Reddy Janga}, {Sabater}, {Sakurikar}, {Seifert}, {Sherbert}, {Sherwood-Taylor}, {Shih}, {Sick}, {Silbiger}, {Singanamalla}, {Singer}, {Sladen}, {Sooley}, {Sornarajah}, {Streicher}, {Teuben}, {Thomas}, {Tremblay}, {Turner}, {Terr{\'o}n}, {van Kerkwijk}, {de la Vega}, {Watkins}, {Weaver}, {Whitmore}, {Woillez}, {Zabalza}, \& {Astropy Contributors}}]{exoplanet:astropy18}
{Astropy Collaboration}, {Price-Whelan}, A.~M., {Sip{\H o}cz}, B.~M., {et~al.} 2018, \aj, 156, 123

\bibitem[{{Astropy Collaboration} {et~al.}(2013){Astropy Collaboration}, {Robitaille}, {Tollerud}, {Greenfield}, {Droettboom}, {Bray}, {Aldcroft}, {Davis}, {Ginsburg}, {Price-Whelan}, {Kerzendorf}, {Conley}, {Crighton}, {Barbary}, {Muna}, {Ferguson}, {Grollier}, {Parikh}, {Nair}, {Unther}, {Deil}, {Woillez}, {Conseil}, {Kramer}, {Turner}, {Singer}, {Fox}, {Weaver}, {Zabalza}, {Edwards}, {Azalee Bostroem}, {Burke}, {Casey}, {Crawford}, {Dencheva}, {Ely}, {Jenness}, {Labrie}, {Lim}, {Pierfederici}, {Pontzen}, {Ptak}, {Refsdal}, {Servillat}, \& {Streicher}}]{exoplanet:astropy13}
{Astropy Collaboration}, {Robitaille}, T.~P., {Tollerud}, E.~J., {et~al.} 2013, \aap, 558, A33

\bibitem[{{Boden} {et~al.}(2005){Boden}, {Sargent}, {Akeson}, {Carpenter}, {Torres}, {Latham}, {Soderblom}, {Nelan}, {Franz}, \& {Wasserman}}]{Boden2005}
{Boden}, A.~F., {Sargent}, A.~I., {Akeson}, R.~L., {et~al.} 2005, \apj, 635, 442

\bibitem[{Bouchy {et~al.}(2025)}]{Bouchy2025}
Bouchy, F. {et~al.} 2025, A\&A, 700, A10

\bibitem[{{Ceppi} {et~al.}(2024){Ceppi}, {Cuello, Nicol\'as}, {Lodato, Giuseppe}, {Longarini, Cristiano}, {Price, Daniel J.}, {Elsender, Daniel}, \& {Bate, Matthew R.}}]{Ceppi2024}
{Ceppi}, S., {Cuello, Nicol\'as}, {Lodato, Giuseppe}, {et~al.} 2024, A\&A, 682, A104

\bibitem[{{Colavita} {et~al.}(2010){Colavita}, {Booth}, {Garcia-Gathright}, {Vasisht}, {Johnson}, \& {Summers}}]{Colavita2010}
{Colavita}, M.~M., {Booth}, A.~J., {Garcia-Gathright}, J.~I., {et~al.} 2010, \pasp, 122, 795

\bibitem[{{Czekala} {et~al.}(2021){Czekala}, {Ribas}, {Cuello}, {Chiang}, {Mac{\'\i}as}, {Duch{\^e}ne}, {Andrews}, \& {Espaillat}}]{Czekala2021}
{Czekala}, I., {Ribas}, {\'A}., {Cuello}, N., {et~al.} 2021, \apj, 912, 6

\bibitem[{{Faruqi} {et~al.}(2025){Faruqi}, {Kennedy}, {Nealon}, \& {Rowther}}]{Faruqi2025}
{Faruqi}, A., {Kennedy}, G., {Nealon}, R., \& {Rowther}, S. 2025, \mnras, 537, 2516

\bibitem[{Foreman-Mackey {et~al.}(2020)Foreman-Mackey, Luger, Czekala, Agol, Price-Whelan, Brandt, Barclay, \& Bouma}]{exoplanet:exoplanet}
Foreman-Mackey, D., Luger, R., Czekala, I., {et~al.} 2020, exoplanet-dev/exoplanet v0.4.0

\bibitem[{{Gallenne} {et~al.}(2015){Gallenne}, {M{\'e}rand}, {Kervella}, {Monnier}, {Schaefer}, {Baron}, {Breitfelder}, {Le Bouquin}, {Roettenbacher}, {Gieren}, {Pietrzy{\'n}ski}, {McAlister}, {ten Brummelaar}, {Sturmann}, {Sturmann}, {Turner}, {Ridgway}, \& {Kraus}}]{Gallenne2015}
{Gallenne}, A., {M{\'e}rand}, A., {Kervella}, P., {et~al.} 2015, \aap, 579, A68

\bibitem[{{Giuppone} \& {Cuello}(2019)}]{GiupponeCuello2019}
{Giuppone}, C.~A. \& {Cuello}, N. 2019, in Journal of Physics Conference Series, Vol. 1365, Journal of Physics Conference Series, 012023

\bibitem[{{Kennedy} {et~al.}(2019){Kennedy}, {Matr{\`a}}, {Facchini}, {Milli}, {Pani{\'c}}, {Price}, {Wilner}, {Wyatt}, \& {Yelverton}}]{Kennedy2019}
{Kennedy}, G.~M., {Matr{\`a}}, L., {Facchini}, S., {et~al.} 2019, Nature Astronomy, 3, 230

\bibitem[{{Kraus} {et~al.}(2020){Kraus}, {Kreplin}, {Young}, {Bate}, {Monnier}, {Harries}, {Avenhaus}, {Kluska}, {Laws}, {Rich}, {Willson}, {Aarnio}, {Adams}, {Andrews}, {Anugu}, {Bae}, {ten Brummelaar}, {Calvet}, {Cur{\'e}}, {Davies}, {Ennis}, {Espaillat}, {Gardner}, {Hartmann}, {Hinkley}, {Labdon}, {Lanthermann}, {LeBouquin}, {Schaefer}, {Setterholm}, {Wilner}, \& {Zhu}}]{Kraus2020}
{Kraus}, S., {Kreplin}, A., {Young}, A.~K., {et~al.} 2020, Science, 369, 1233

\bibitem[{{Le Bouquin} {et~al.}(2011){Le Bouquin}, {Berger}, {Lazareff}, {Zins}, {Haguenauer}, {Jocou}, {Kern}, {Millan-Gabet}, {Traub}, {Absil}, {Augereau}, {Benisty}, {Blind}, {Bonfils}, {Bourget}, {Delboulbe}, {Feautrier}, {Germain}, {Gitton}, {Gillier}, {Kiekebusch}, {Kluska}, {Knudstrup}, {Labeye}, {Lizon}, {Monin}, {Magnard}, {Malbet}, {Maurel}, {M{\'e}nard}, {Micallef}, {Michaud}, {Montagnier}, {Morel}, {Moulin}, {Perraut}, {Popovic}, {Rabou}, {Rochat}, {Rojas}, {Roussel}, {Roux}, {Stadler}, {Stefl}, {Tatulli}, \& {Ventura}}]{LeBouquin2011}
{Le Bouquin}, J.~B., {Berger}, J.~P., {Lazareff}, B., {et~al.} 2011, \aap, 535, A67

\bibitem[{{Mason} {et~al.}(2001){Mason}, {Wycoff}, {Hartkopf}, {Douglass}, \& {Worley}}]{WDS2001}
{Mason}, B.~D., {Wycoff}, G.~L., {Hartkopf}, W.~I., {Douglass}, G.~G., \& {Worley}, C.~E. 2001, \aj, 122, 3466

\bibitem[{Mayor {et~al.}(2003)}]{Mayor2003}
Mayor, M. {et~al.} 2003, The Messenger, 114, 20

\bibitem[{{Mennesson} {et~al.}(2014){Mennesson}, {Millan-Gabet}, {Serabyn}, {Colavita}, {Absil}, {Bryden}, {Wyatt}, {Danchi}, {Defr{\`e}re}, {Dor{\'e}}, {Hinz}, {Kuchner}, {Ragland}, {Scott}, {Stapelfeldt}, {Traub}, \& {Woillez}}]{Mennesson2014}
{Mennesson}, B., {Millan-Gabet}, R., {Serabyn}, E., {et~al.} 2014, \apj, 797, 119

\bibitem[{{Merle} {et~al.}(2024){Merle}, {Pourbaix, D.}, {Jorissen, A.}, {Siopis, C.}, {Van Eck, S.}, \& {Van Winckel, H.}}]{Merle2024}
{Merle}, T., {Pourbaix, D.}, {Jorissen, A.}, {et~al.} 2024, A\&A, 684, A74

\bibitem[{Méndez {et~al.}(2025)Méndez, Tokovinin, Costa, \& Dirk}]{Mendez2025}
Méndez, R.~A., Tokovinin, A., Costa, E., \& Dirk, M. 2025, AJ, 169, 226

\bibitem[{{Ribas} {et~al.}(2026){Ribas}, {Lack}, {Zagaria}, {Mac{\'\i}as}, {Andrews}, {Bayo}, {Clarke}, {Cuello}, \& {Espaillat}}]{Ribas2026}
{Ribas}, {\'A}., {Lack}, T., {Zagaria}, F., {et~al.} 2026, \mnras, 548, stag641

\bibitem[{{Ribas} {et~al.}(2018){Ribas}, {Mac{\'\i}as}, {Espaillat}, \& {Duch{\^e}ne}}]{Ribas2018}
{Ribas}, {\'A}., {Mac{\'\i}as}, E., {Espaillat}, C.~C., \& {Duch{\^e}ne}, G. 2018, \apj, 865, 77

\bibitem[{{Ronco} {et~al.}(2021){Ronco}, {Guilera}, {Cuadra}, {Miller Bertolami}, {Cuello}, {Fontecilla}, {Poblete}, \& {Bayo}}]{Ronco2021}
{Ronco}, M.~P., {Guilera}, O.~M., {Cuadra}, J., {et~al.} 2021, \apj, 916, 113

\bibitem[{Salvatier {et~al.}(2016)Salvatier, Wiecki, \& Fonnesbeck}]{exoplanet:pymc3}
Salvatier, J., Wiecki, T.~V., \& Fonnesbeck, C. 2016, PeerJ Computer Science, 2, e55

\bibitem[{{Scott} {et~al.}(2021){Scott}, {Howell}, {Gnilka}, {Stephens}, {Salinas}, {Matson}, {Furlan}, {Horch}, {Everett}, {Ciardi}, {Mills}, \& {Quigley}}]{Scott2021}
{Scott}, N.~J., {Howell}, S.~B., {Gnilka}, C.~L., {et~al.} 2021, Frontiers in Astronomy and Space Sciences, 8, 138

\bibitem[{{Soderblom} {et~al.}(1998){Soderblom}, {King}, {Siess}, {Noll}, {Gilmore}, {Henry}, {Nelan}, {Burrows}, {Brown}, {Perryman}, {Benedict}, {McArthur}, {Franz}, {Wasserman}, {Jones}, {Latham}, {Torres}, \& {Stefanik}}]{Soderblom1998}
{Soderblom}, D.~R., {King}, J.~R., {Siess}, L., {et~al.} 1998, \apj, 498, 385

\bibitem[{{The Theano Development Team} {et~al.}(2016){The Theano Development Team}, {Al-Rfou}, {Alain}, {Almahairi}, {Angermueller}, {Bahdanau}, {Ballas}, {Bastien}, {Bayer}, {Belikov}, {Belopolsky}, {Bengio}, {Bergeron}, {Bergstra}, {Bisson}, {Bleecher Snyder}, {Bouchard}, {Boulanger-Lewandowski}, {Bouthillier}, {de Br{\'e}bisson}, {Breuleux}, {Carrier}, {Cho}, {Chorowski}, {Christiano}, {Cooijmans}, {C{\^o}t{\'e}}, {C{\^o}t{\'e}}, {Courville}, {Dauphin}, {Delalleau}, {Demouth}, {Desjardins}, {Dieleman}, {Dinh}, {Ducoffe}, {Dumoulin}, {Ebrahimi Kahou}, {Erhan}, {Fan}, {Firat}, {Germain}, {Glorot}, {Goodfellow}, {Graham}, {Gulcehre}, {Hamel}, {Harlouchet}, {Heng}, {Hidasi}, {Honari}, {Jain}, {Jean}, {Jia}, {Korobov}, {Kulkarni}, {Lamb}, {Lamblin}, {Larsen}, {Laurent}, {Lee}, {Lefrancois}, {Lemieux}, {L{\'e}onard}, {Lin}, {Livezey}, {Lorenz}, {Lowin}, {Ma}, {Manzagol}, {Mastropietro}, {McGibbon}, {Memisevic}, {van Merri{\"e}nboer}, {Michalski}, {Mirza}, {Orlandi}, {Pal}, {Pascanu}, {Pezeshki}, {Raffel},
  {Renshaw}, {Rocklin}, {Romero}, {Roth}, {Sadowski}, {Salvatier}, {Savard}, {Schl{\"u}ter}, {Schulman}, {Schwartz}, {Vlad Serban}, {Serdyuk}, {Shabanian}, {Simon}, {Spieckermann}, {Ramana Subramanyam}, {Sygnowski}, {Tanguay}, {van Tulder}, {Turian}, {Urban}, {Vincent}, {Visin}, {de Vries}, {Warde-Farley}, {Webb}, {Willson}, {Xu}, {Xue}, {Yao}, {Zhang}, \& {Zhang}}]{exoplanet:theano}
{The Theano Development Team}, {Al-Rfou}, R., {Alain}, G., {et~al.} 2016, arXiv e-prints, arXiv:1605.02688

\bibitem[{Thompson {et~al.}(2017)Thompson, Moran, \& Swenson}]{Thompson2017}
Thompson, A.~R., Moran, J.~M., \& Swenson, G.~W. 2017, Interferometry and Synthesis in Radio Astronomy, 3rd edn. (Springer)

\bibitem[{{Tokovinin}(2018)}]{Tokovinin2018a}
{Tokovinin}, A. 2018, \apjs, 235, 6

\bibitem[{{Torres} {et~al.}(2006){Torres}, {Quast, G. R.}, {da Silva, L.}, {de la Reza, R.}, {Melo, C. H. F.}, \& {Sterzik, M.}}]{Torres2006}
{Torres}, C. A.~O., {Quast, G. R.}, {da Silva, L.}, {et~al.} 2006, A\&A, 460, 695

\bibitem[{Torres {et~al.}(1999)Torres, Henry, Franz, \& Wasserman}]{Torres1999}
Torres, G., Henry, T.~J., Franz, O.~G., \& Wasserman, L.~H. 1999, AJ, 117, 562

\bibitem[{{Z{\'u}{\~n}iga-Fern{\'a}ndez} {et~al.}(2021){Z{\'u}{\~n}iga-Fern{\'a}ndez}, {Bayo}, {Elliott}, {Zamora}, {Corval{\'a}n}, {Haubois}, {Corral-Santana}, {Olofsson}, {Hu{\'e}lamo}, {Sterzik}, {Torres}, {Quast}, \& {Melo}}]{Zuniga-Fernandez2021}
{Z{\'u}{\~n}iga-Fern{\'a}ndez}, S., {Bayo}, A., {Elliott}, P., {et~al.} 2021, \aap, 645, A30

\bibitem[{{Z\'u\~niga-Fern\'andez} {et~al.}(2021){Z\'u\~niga-Fern\'andez}, {Olofsson}, {Bayo}, {Haubois, X.}, {Corral-Santana, J. M.}, {Lopera-Mej\'{\i}a, A.}, {Ronco, M. P.}, {Tokovinin, A.}, {Gallenne, A.}, {Kennedy, G. M.}, \& {Berger, J.-P.}}]{SZF2021}
{Z\'u\~niga-Fern\'andez}, S., {Olofsson}, J., {Bayo}, A., {et~al.} 2021, A\&A, 655, A15

\end{thebibliography}

\begin{appendix}
\section{Observations}\label{sec:appendix_observations}

This section presents complementary information regarding the observations used in this work.

\paragraph{PIONIER} For PIONIER observation, we determined the astrometric positions by fitting the $V^2$ and CP with a binary model using the interferometric tool \texttt{GUIcandid}\footnote{\url{https://github.com/agallenne/GUIcandid}} \citep{Gallenne2015}. For the new BaBb astrometric data see Table \ref{tab:astrometry}.

\paragraph{NIRPS/HARPS RVs} Most of the RV measurements used in this work were published by \cite{SZF2021} and \cite{Merle2024}. Here we present the new RV measurements from NIRPS/HARPS observations, see Table \ref{tab:appendix_RV_AaAb} and Table \ref{tab:appendix_RV_BaBb}. The RVs were measured from the CCFs of the spectra using a CORAVEL-type M2 numerical mask using a standalone CCF tool\footnote{\url{https://github.com/szunigaf/CCF_functions}}
(for further details, see \citealt{Zuniga-Fernandez2021}).

\paragraph{HD98800\,AB RVs} The radial velocities used to fit the AB orbit were derived from the systemic velocity at the mean epoch of each data set and are listed in Table~\ref{tab:appendix_RV_AB}.

\begin{table}[h]
\small
\centering
\caption{Relative astrometric position of the BaBb secondary component, flux ratio, and resolved flux from PIONIER data.}
\label{tab:astrometry}
\begin{tabular}{l c}
\toprule
Parameters &  \\
\midrule
MJD & 59640.213272 \\
$\Delta \alpha$ (mas) & 16.4982 \\
$\Delta \delta$ (mas) & -13.8899 \\
$\sigma_\mathrm{PA}$ ($^\circ$) & 152.5985 \\
$\sigma_\mathrm{maj}$ (mas) & 0.0114 \\
$\sigma_\mathrm{min}$ (mas) & 0.0099 \\
$f_{~\mathrm{Bb}}/f_{~\mathrm{Ba}}$ (\%) & $61 \pm 1$ \\
$f_{~\mathrm{res}}$\tablefootmark{a} (\%) & $76 \pm 2$ \\
Baselines & K0-G2-D0-J3 \\
Seeing (\arcsec) & 0.83 \\
$\tau_0$ (ms) & 4.12 \\
\bottomrule
\end{tabular}
\tablefoot{\tablefoottext{a}{Parameter to take the background cross-contamination into account (non-coherent light), parametrised in \texttt{CANDID} as a resolved flux. The last two rows correspond to the atmospheric conditions: the seeing and coherence time ($\tau_0$), measured by the seeing monitor.}}
\end{table}

\begin{table}[H]
\small
                \centering
                \caption{Radial velocity measurements for AaAb subsystem.}
                \begin{tabular}{lccr}
                        \toprule
                        MJD &  $~\mathrm{RV_{~\mathrm{Aa}}}$  & $~\mathrm{\sigma_{~\mathrm{Aa}}}$ & Instrument \\
                        & (km s$^{-1}$) & (km s$^{-1}$) &\\
                        \midrule
                        60097.996254 & 18.00 & 0.05 & HARPS \\
                        60107.955827 & 19.431 & 0.077 & HARPS \\
                        60107.970162 & 19.785 & 0.088 & HARPS \\
                        60128.935957 & 9.529  & 0.175 & HARPS \\
                        60097.996254 & 20.31  & 0.10 & NIRPS \\
                        60107.955827 & 18.849 & 0.837 & NIRPS \\
                        60107.970162 & 18.251 & 0.733 & NIRPS \\
                        60128.935957 & 7.872  & 0.521 & NIRPS \\
            \bottomrule
                \end{tabular}
                \label{tab:appendix_RV_AaAb}
        \end{table}

        \begin{table}[H]
\small
                \centering
                \caption{Radial velocity measurements for BaBb subsystem.}
                \begin{tabular}{lcccr}
                        \toprule
                        MJD &  $~\mathrm{RV_{~\mathrm{Ba}}}$  & $~\mathrm{\sigma_{~\mathrm{Ba}}}$ & Instrument \\
                        & (km s$^{-1}$) & (km s$^{-1}$) &\\
                        \midrule
                        60097.996254 & --2.28 & 0.08 & HARPS \\
                        60107.955827 & --1.654 & 0.069 & HARPS \\
                        60107.970162 & --1.534 & 0.069 & HARPS \\
                        660128.935957 & 1.221 & 0.069 & HARPS \\
                        60097.996254 & --3.42 & 0.11 & NIRPS \\
                        60107.955827 & --1.829 & 0.123 & NIRPS \\
                        60107.970162 & --1.711 & 0.156 & NIRPS \\
                        60128.935957 & 1.28 & 0.094 & NIRPS \\
            \midrule
                        MJD &  $~\mathrm{RV_{~\mathrm{Bb}}}$  & $~\mathrm{\sigma_{~\mathrm{Bb}}}$  & Instrument \\
                        & (km s$^{-1}$)  & (km s$^{-1}$) &\\
                         \midrule
            60097.996254 & 13.68 & 2.09 & HARPS \\
            60107.955827 & 17.009 & 0.108 & HARPS \\
            60107.970162 & 17.359 & 0.071 & HARPS \\
            60128.935957 & 14.346 & 0.127 & HARPS \\
            60097.996254 & 18.72 & 2.76 & NIRPS \\
            60107.955827 & 18.387 & 0.616 & NIRPS \\
            60107.970162 & 18.534 & 0.939 & NIRPS\\
            60128.935957 & 14.285 & 0.126 & NIRPS \\
            \bottomrule
                \end{tabular}
                \label{tab:appendix_RV_BaBb}
        \end{table} 

                \begin{table}[H]
\small 
                \centering
                \caption{Radial velocity measurements for AB system.}
                \begin{tabular}{lccr}
                        \toprule
                        Median MJD &  $~\mathrm{RV_{~\mathrm{AaAb}}}$ & $~\mathrm{\sigma_{~\mathrm{AaAb}}}$ & Source \\
                        & (km s$^{-1}$) & (km s$^{-1}$)  & \\
                          \midrule
            48635.4564 & 12.8 & 0.1  & TO95 \\
            50841.6274 & 12.1 & 0.5  & ELODIE \\
            54311.9669 & 14.7 & 0.4  & FEROS \tablefootmark{a}\\
            57062.7727 & 12 & 2  & FEROS \tablefootmark{b}\\
            59375.5439 & 11.8 & 0.2  & CTIO \\
            57054.6979 & 12.9 & 0.2  & Mercator \\
            60107.9629 & 11.8 & 0.8  & HARPS \\
            60107.9629 & 11.6 & 0.7  & NIRPS \\
            \midrule
            Median MJD &  $~\mathrm{RV_{~\mathrm{BaBb}}}$ & $~\mathrm{\sigma_{~\mathrm{BaBb}}}$ & Source \\
                        & (km s$^{-1}$)  & (km s$^{-1}$)  & \\
                        \midrule
                        48635.4564 &  5.6 & 0.1  & TO95 \\
            50841.6274 & 3.4 & 0.7 & ELODIE \\
            58072.3724 & 5.1 & 1  & KE19 \tablefootmark{c} \\
            59375.5439 & 6.4 & 0.4  & CTIO \\
            57399.7161 & 4.6 & 0.3  & Mercator \\
            60107.9629 & 7.1 & 0.1  & HARPS \\
            60107.9629 & 7.0 & 0.1 & NIRPS \\
            \bottomrule
                \end{tabular}
                \tablefoot{\tablefoottext{a}{From FEROS observations taken in 2007.}\tablefoottext{b}{From FEROS observation taken in 2015.}\tablefoottext{c}{\cite{Kennedy2019}.}}
                \label{tab:appendix_RV_AB}
                \vspace{-0.5cm}
        \end{table} 

\begin{table}[h]
\small 
                \centering
                \caption{VLA astrometry measurements of AB system.}
                \begin{tabular}{lcccc}
                        \toprule
                        Date &  $\rho$  & $\sigma_{\rho}$ & $\theta$ & $\sigma_{\theta}$ \\
                        & (\arcsec) & (\arcsec) & ($^\circ$) & ($^\circ$)  \\
                        \midrule
                        2011.5097 \tablefootmark{a} & 0.614 & 0.088 & 3.07 & 0.82 \\
                        2018.22 & 0.431 & 0.112 & 22.64 & 3.6 \\
            \bottomrule
                \end{tabular} 
                \label{tab:appendix_ABastrometry}
                \tablefoot{\tablefoottext{a}{From C-band observations presented by \cite{Ribas2018}.}}
                \vspace{-0.2cm}
        \end{table}

\paragraph{HD98800\,AB astrometry} The AB astrometric measurements before 2016 are available in the Washington Double Star Catalogue \citep[WDS][]{WDS2001} and \cite{Tokovinin2018a}. The astrometric measurements from speckle interferometry at SOAR up to 2021 were presented by \cite{SZF2021}, and those obtained with the Zorro camera on Gemini South are available in \cite{Mendez2025}. The new VLA observations were presented by \cite{Ribas2026}, who resolved the disc around HD~98800~B at 6.8~mm and detected emission from both the A and B components at 3~cm. In this work, we use the latter data set since it detects both components, allowing for relative astrometry, and the lower angular resolution at 3~cm means that the individual components are not resolved internally, which facilitates centroid determination. The observations were taken on 23 May 2018, and the resulting synthesized beam was $0.37\arcsec \times 0.17\arcsec$. The position of each source was determined by fitting two Gaussians in the image plane using the \texttt{imfit} task in \texttt{CASA}. The VLA astrometric measurements are listed in Table~\ref{tab:appendix_ABastrometry}.

\section{Disc-plane projection and crossing-time estimate}\label{sec:app_occultation}

To estimate the epochs at which AaAb crosses the projected BaBb disc, we evaluated the posterior orbital motion of A relative to B directly in the sky plane. For each orbital solution, we used the posterior percentile tracks of the sky-plane separation $\rho(t)$ and position angle $\theta(t)$ of A relative to B, and compared them with the sky-plane projection of the adopted disc geometry. Assuming that the disc is intrinsically circular, its projection on the sky is an ellipse whose semiminor axis is reduced by a factor of $\cos i$, where $i$ is the disc inclination. We adopted $i=26^\circ$ and ${\rm PA}=15.6^\circ$, together with dust radii of 2.5 and 4.6 au and gas radii of 1.6 and 6.4 au from \citet{Kennedy2019} and \citet{Faruqi2025}, and converted these radii to angular units using $d=45$ pc. The disc edges were then projected onto the sky plane as ellipses and compared with the sky-plane orbital tracks. Crossing times were identified from sign changes between the orbital position and each projected disc boundary, followed by linear interpolation between adjacent time samples. Repeating this procedure for the 16th, 50th, and 84th percentile tracks yields the median ingress and egress epochs and their formal 1$\sigma$ intervals.

\section{Orbital fitting complementary information}\label{sec:apendix_orbits}
\begin{figure}[h]
   \centering
   \includegraphics[width=0.48\textwidth]{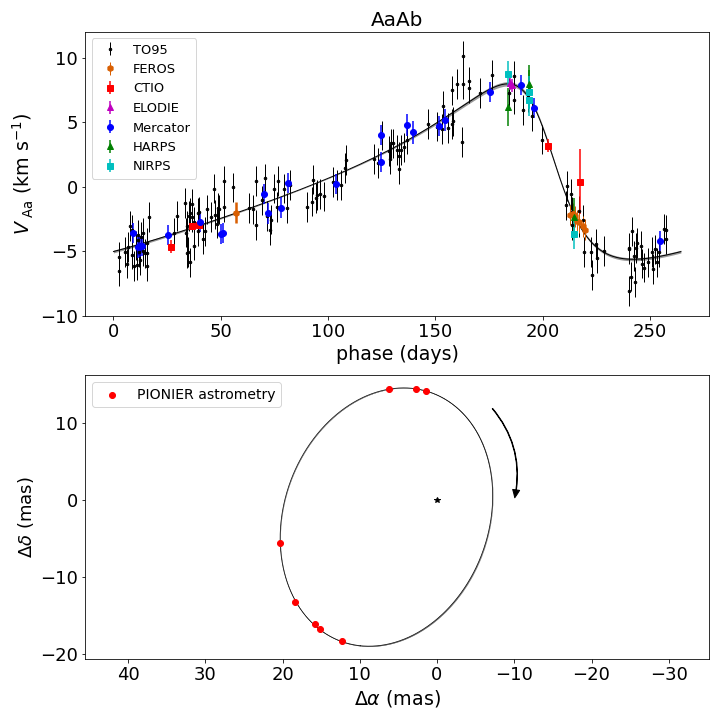}
   \vspace{-0.25cm}
   \caption{Best orbital solution for AaAb. In both panels, the solid line corresponds to the best-fit model. \textit{Bottom panel:} The primary star Aa is located at the origin. The relative positions of Ab are plotted as filled dots; the error ellipses from PIONIER astrometry are smaller than the marker. \textit{Upper panel:} The coloured markers correspond to the primary star RV measurements. The systematic velocity $\gamma$ for each set of observations was subtracted.} 
  \label{Fig:AaAb_orbit}%
    \end{figure}

\begin{figure}[h]
\centering

\begin{subfigure}{0.49\textwidth}
    \centering
    \includegraphics[width=\textwidth]{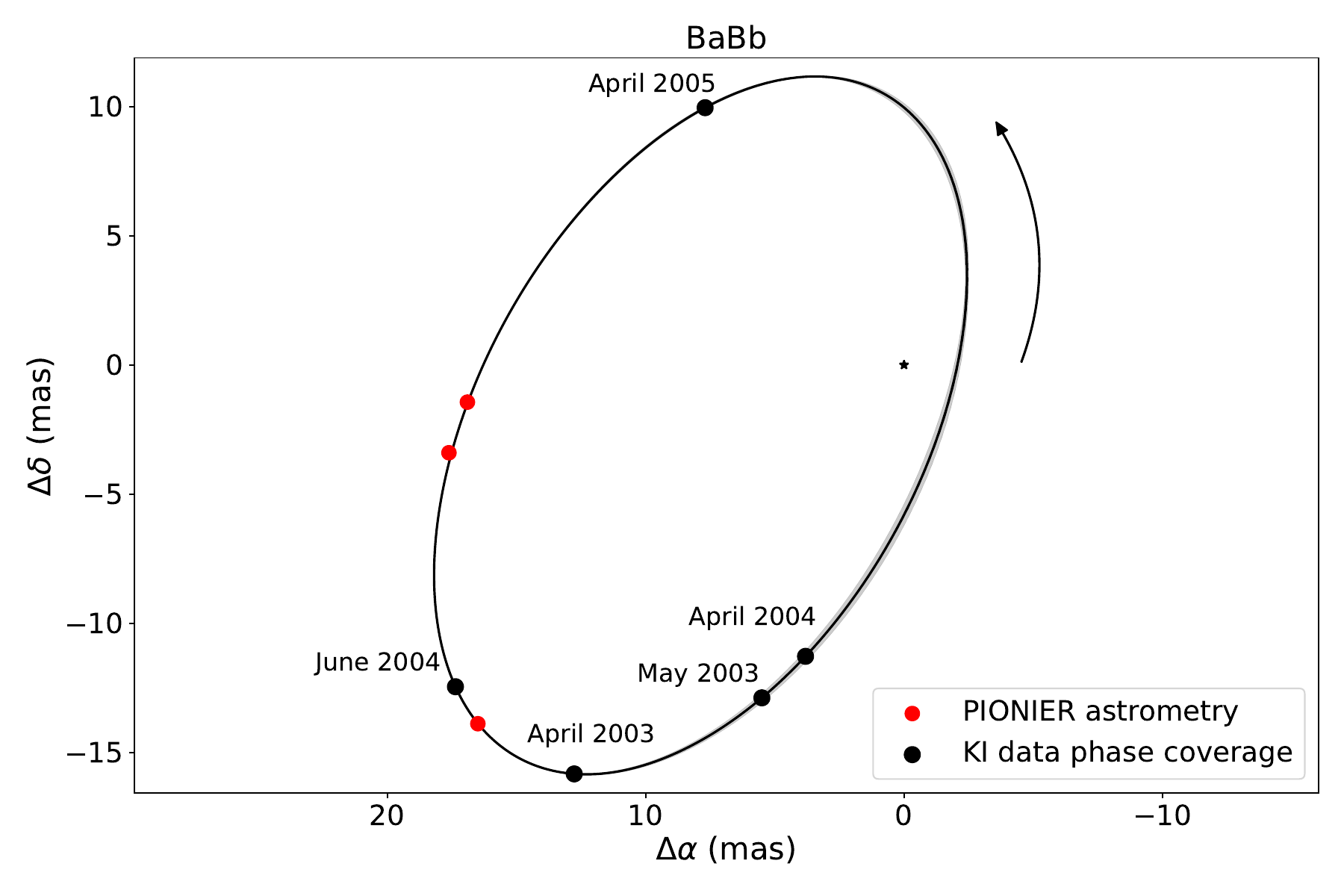}
    \label{Fig:BaBb_orbit}
\end{subfigure}

\begin{subfigure}{0.49\textwidth}
    \centering
    \includegraphics[width=\textwidth]{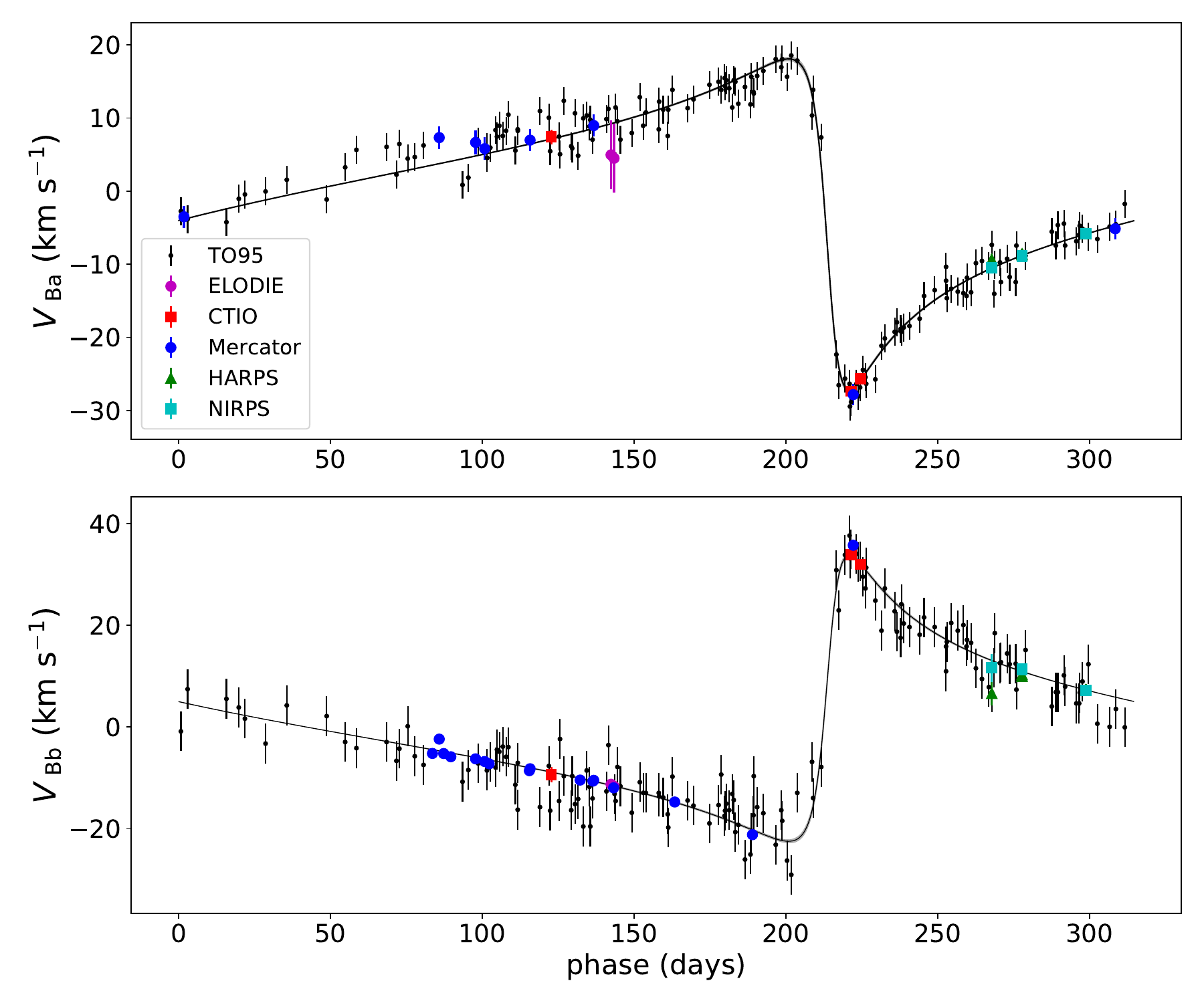}
    \label{Fig:BaBb_RVorbit}
\end{subfigure}
\vspace{-0.25cm}
\caption{Best orbital solution for BaBb. In both panels, the solid line corresponds to the best-fit model. \textit{Top panel:} The shaded area to the $1\sigma$ region. The primary star Ba is located at the origin. The relative positions of Bb are plotted as filled dots. The error ellipses from PIONIER astrometry are smaller than the markers. \textit{Bottom panel:} Phase-folded RVs for BaBb. The systemic velocity $\gamma$ for each set of observations was subtracted.}
\label{Fig:BaBb_combined}
\end{figure}

\begin{figure}[h]
   \centering
   \includegraphics[width=0.41\textwidth]{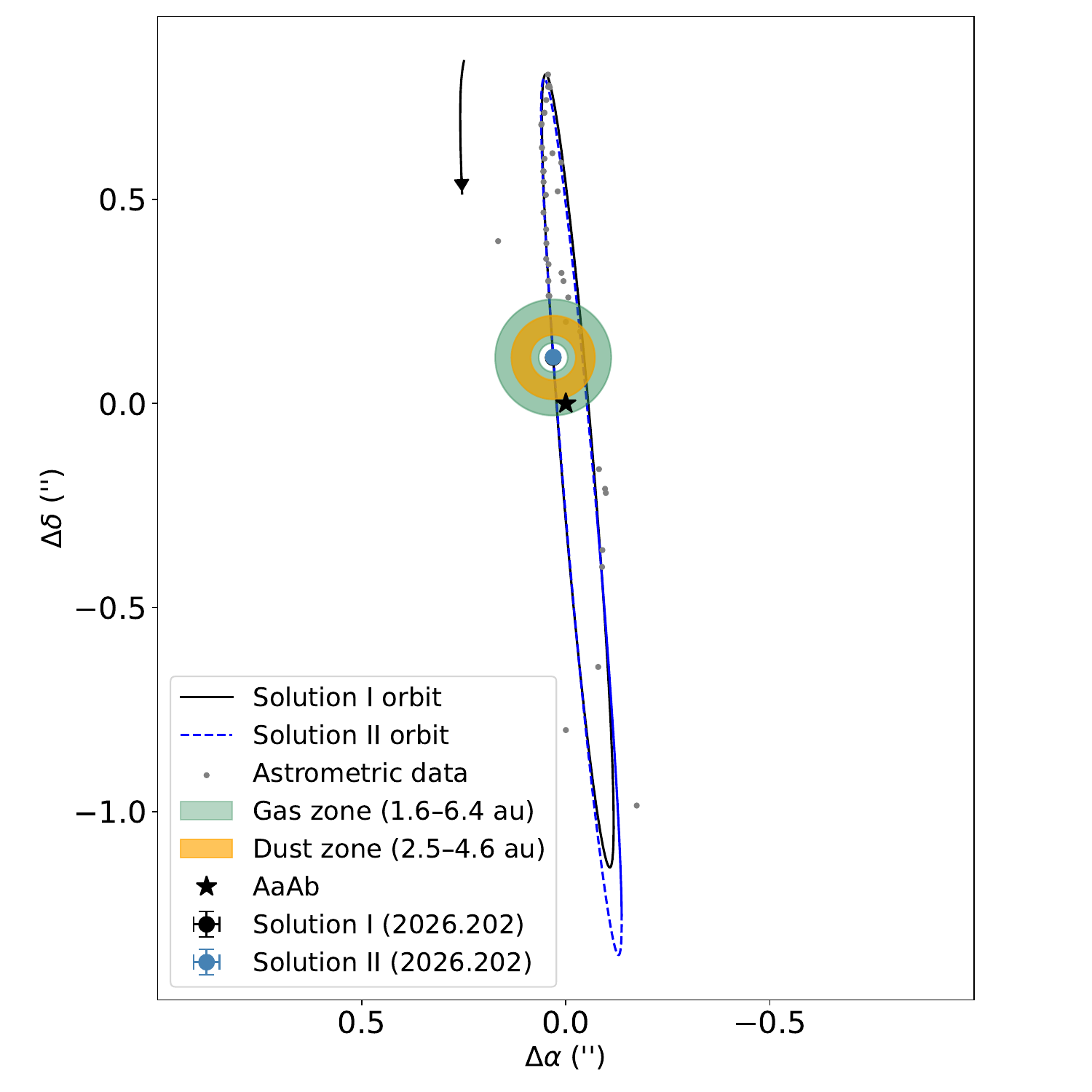}
   \caption{Best orbital solution for the AB outer orbit for the two assumed astrometric uncertainty values for data take before 1991. The solid black line corresponds to Solution I, and the dashed blue line to Solution II. The grey dots represent the astrometric measurements. A sketch of the mid-March disc position in the sky plane, based on the disc geometry adopted by \cite{Faruqi2025}, is overplotted for both solutions.} 
\label{Fig:AB_disc_model}
    \end{figure}

\begin{figure}[h]
\centering
\includegraphics[width=0.4\textwidth]{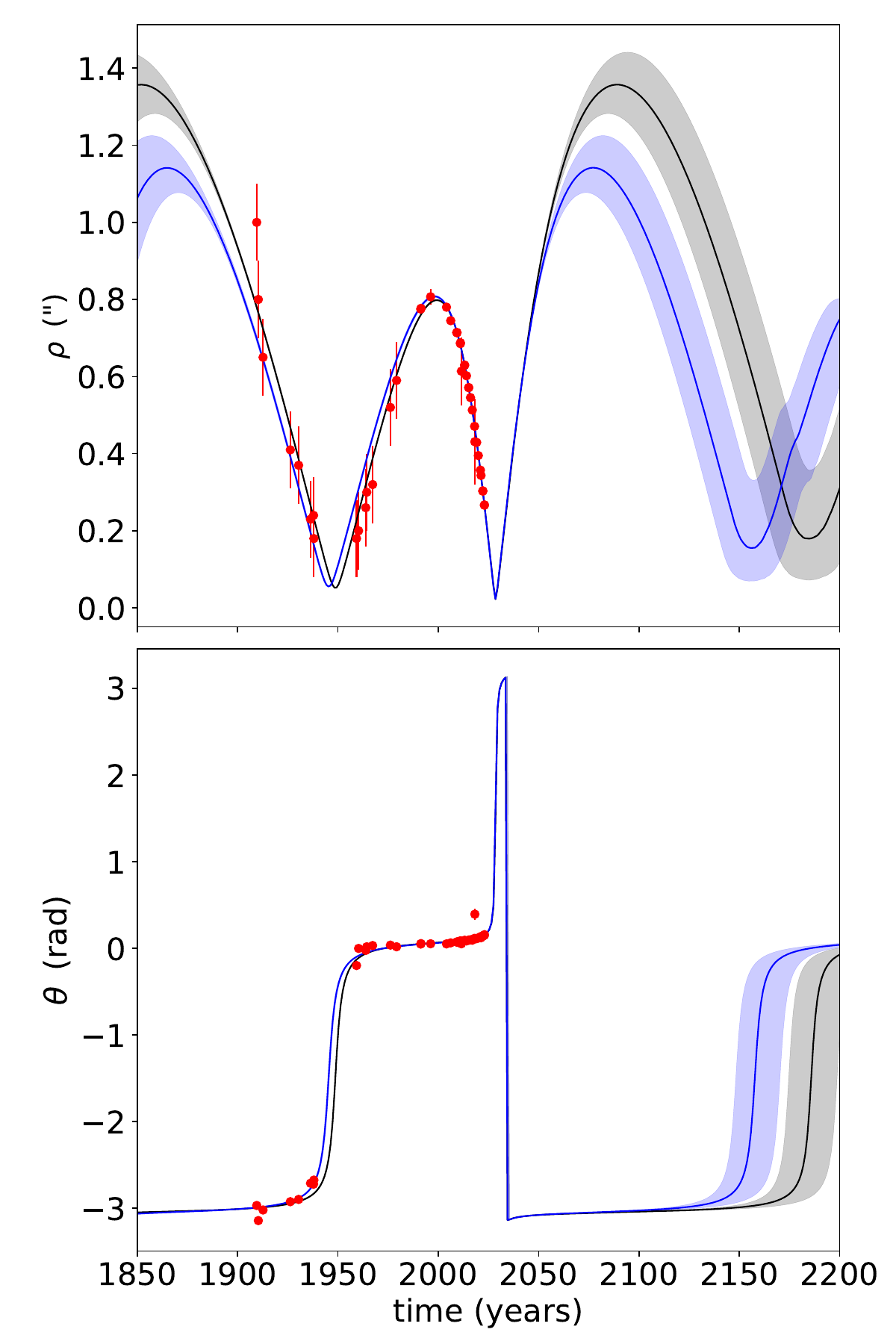}
\vspace{-0.2cm}
\caption{Best orbital solution for AB orbit. In all panels the solid line corresponds to the best-fit model and the shaded area to the $1\sigma$ region. The solid blue lines correspond to Solution I and the black ones to Solution II. The red dots correspond to the astrometric measurements. The error bars shown in the astrometry before 1991 correspond to the large uncertainty case.}
\label{Fig:AB_orbit_sepPA}
\end{figure}

\begin{figure*}[h]
\centering
\begin{subfigure}{\textwidth}
\centering
\includegraphics[width=0.9\linewidth]{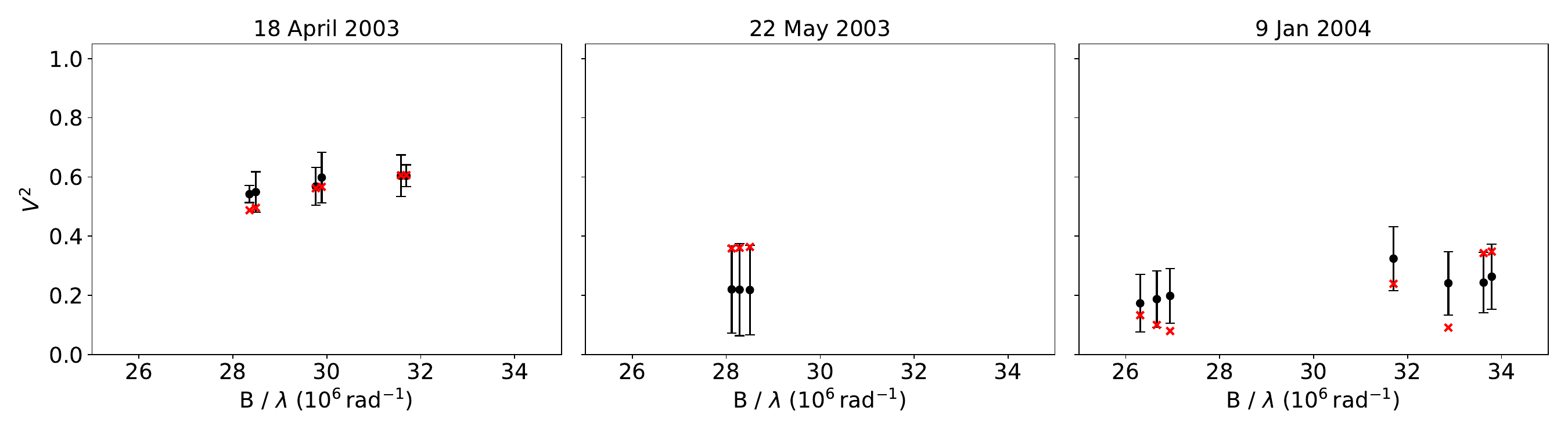}
\end{subfigure}

\begin{subfigure}{\textwidth}
\centering
\includegraphics[width=0.9\linewidth]{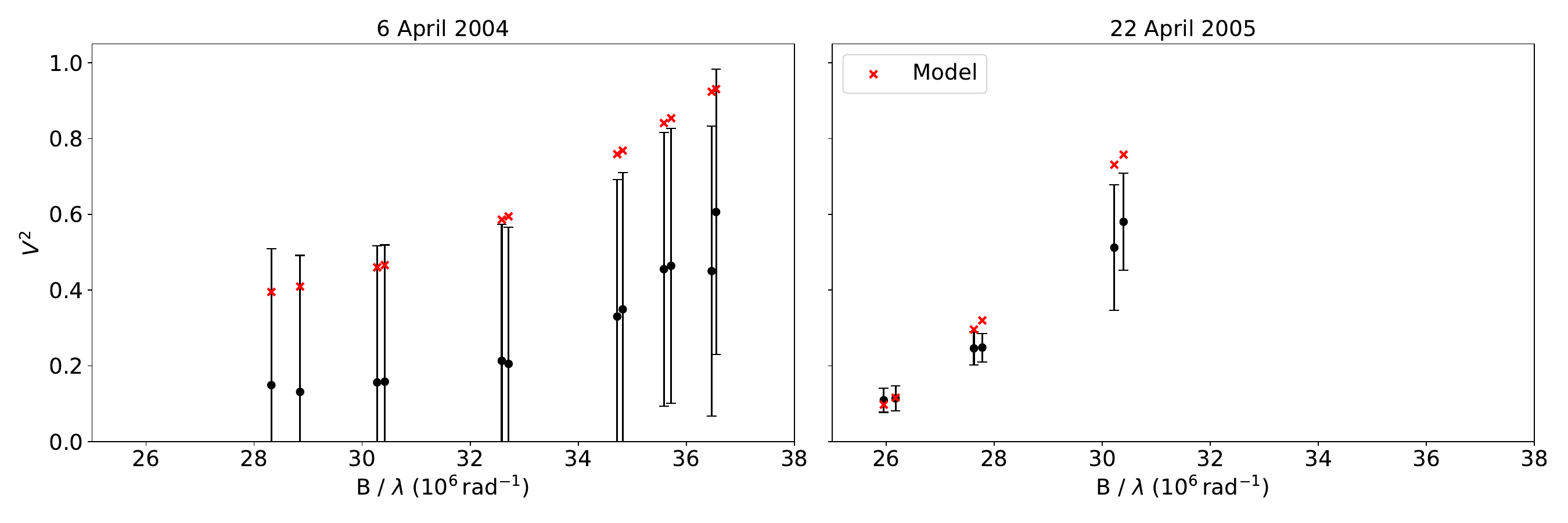}
\end{subfigure}

\caption{Squared visibilities from KI observations published in \cite{Boden2005}. The black circles represent the observed values and the red crosses represent the best-fit BaBb binary model from this work. The plotted uncertainties include an extra jitter term, added in quadrature to the pipeline $V^2$ errors, to account for instrumental systematics and atmospheric phase fluctuations (see Section \ref{sec:orbital_fit}).}
\label{fig:KI_model}
\end{figure*}

\begin{figure}[h]
\centering
\includegraphics[width=0.49\textwidth]{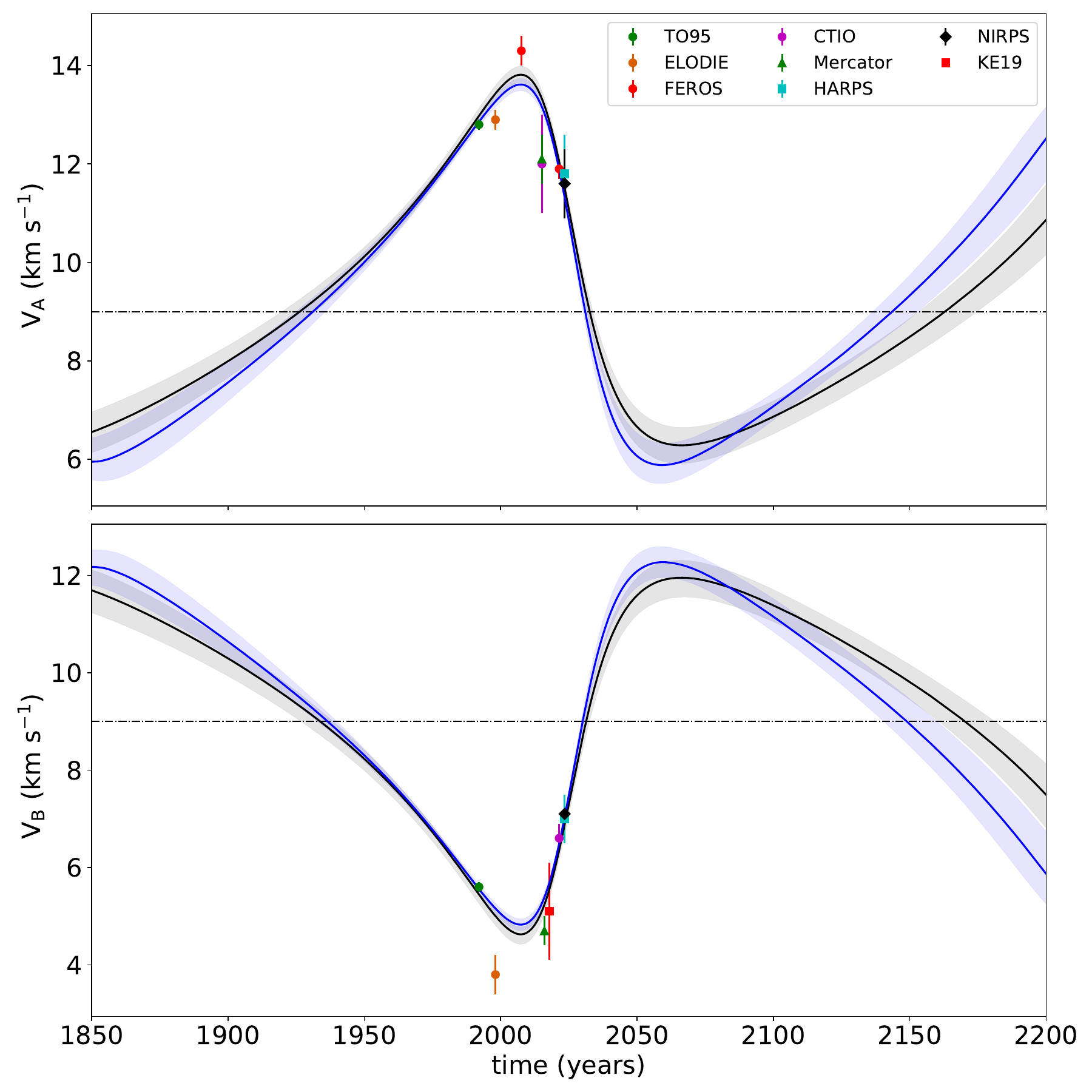}
\vspace{-0.3cm}
\caption{Best orbital solution for AB outer orbit. In both panels the solid line corresponds to the best fit model and the shaded area to the $1\sigma$ region. The solid blue lines correspond to Solution I and the black ones to Solution II.  The dots markers correspond to the RV measurement of systemic velocities from our orbital solutions.}
\label{Fig:AB_orbit_RV}
\end{figure}

This section presents the prior distributions used for each orbital fitting. Additionally, we also show the best orbital solution from the posterior samples of each MCMC model (see Fig. \ref{Fig:AaAb_orbit} to \ref{Fig:AB_orbit_RV}). 

\begin{table}[h]
\small
                \centering
                \caption{Prior distribution used in HD98800\,AB orbital fitting.}
                \begin{tabular}{lr}
                        \toprule
                        Parameters &  AB\\
                        \midrule
                        Period (years)       & $\log \mathcal{U}$ $[100,\,500]$ \\
            T$_{0}$ (yr)       & $\mathcal{U}$ $[2\,000,\,2\,040]$ \\
            $e$                 & $\mathcal{U}$ $[0,\,1]$  \\
            $\omega_{~\mathrm{AaAb}}$ ($^\circ$) & $\mathcal{U}$ $[0,\,360]$  \\
            $\Omega$ ($^\circ$)      & $\mathcal{U}$ $[0,\,360]$   \\
            $~\mathrm{cos\,(i)}$ & $\mathcal{U}$ $[-1,\,1]$    \\
            $M_{~\mathrm{A}}$ (M$_{\sun}$) & $\mathcal{N}$ $[1.25,\,0.1]$ \\
            $M_{~\mathrm{B}}$ (M$_{\sun}$) & $\mathcal{N}$ $[1.37,\,0.1]$ \\
            $\pi$  (mas)     & $\mathcal{N}$ $[22.6,\,0.4]$ \\
            $\gamma_{~\mathrm{AB}}$ (km s$^{-1}$) & $\mathcal{U}$ $[0,\,20]$\\
            \bottomrule 
                \end{tabular}
                \label{tab:appendix_priorOuter}
        \end{table}

\begin{table}[h]
\small
                \centering
                \caption{Prior distribution used in HD98800\,AaAb and BaBb orbital fitting.}
                \begin{tabular}{lrr}
                        \toprule
                        Parameters &  AaAb & BaBb \\
                         \midrule
                        Period (d)       & $\log \mathcal{U}$ $[200,\,300]$  & $\log \mathcal{U}$ $[250,\,350]$  \\
            T$_{0}$ (MJD)       & $\mathcal{N}$ $[48\,737,\,20]$ & $\mathcal{N}$ $[48\,709,\,20]$ \\
            $e$                 & $\mathcal{U}$ $[0,\,1]$  & $\mathcal{U}$ $[0,\,1]$ \\
            $\omega_{~\mathrm{primary}}$ ($^\circ$) & $\mathcal{U}$ $[0,\,360]$ & $\mathcal{U}$ $[0,\,360]$ \\
            $\Omega$ ($^\circ$)      & $\mathcal{U}$ $[0,\,360]$  & $\mathcal{U}$ $[0,\,360]$ \\
            $\cos{i}$ & $\mathcal{U}$ $[-1,\,1]$  & $\mathcal{U}$ $[-1,\,1]$  \\
            $a$ (mas)            & $\mathcal{U}$ $[5,\,30]$  & $\mathcal{U}$ $[5,\,30]$  \\
            $K_{1}$ (km s$^{-1}$)        & $\mathcal{U}$ $[0,\,20]$ & $\mathcal{U}$ $[0,\,50]$ \\
            $K_{2}$  (km s$^{-1}$)       & \dots  & $\mathcal{U}$ $[0,\,50]$  \\
            $\gamma$  (km s$^{-1}$)     & $\mathcal{U}$ $[0,\,20]$ & $\mathcal{U}$ $[0,\,20]$ \\
            \bottomrule 
                \end{tabular}
                \label{tab:appendix_priorInner}
        \end{table}


\end{appendix}

\end{document}